# Multi-Dimensional Geometric Complexity in Urban Transportation Systems

**Farideddin Peiravian[a,1] and Sybil Derrible[a]**

[a] University of Illinois at Chicago, Complex and Sustainable Urban Networks (CSUN) Lab, Civil Engineering Department, 842 W. Taylor St. (MC246), Chicago, IL, US 60607

[1] Email address: fpeira2@uic.edu, Tel: 312-996-2429

**Abstract**

Transportation networks serve as windows into the complex world of urban systems. By properly characterizing a road network, we can therefore better understand its encompassing urban system. This study offers a geometrical approach towards capturing inherent properties of urban road networks. It offers a robust and efficient methodology towards defining and extracting three relevant indicators of road networks: area, line, and point thresholds, through measures of their grid equivalents. By applying the methodology to 50 U.S. urban systems, we successfully observe differences between eastern versus western, coastal versus inland, and old versus young, cities. Moreover, we show that many socio-economic characteristics as well as travel patterns within urban systems are directly correlated with their corresponding area, line, and point thresholds.

**Significance Statement**

Cities are complex systems, consisting of many interrelated components that have evolved over long periods of time. In particular, their transportation systems physically exhibit this evolutionary process. Much information can therefore be collected by studying the geometry of urban transportation systems. This study offers a simple yet robust method to capture three geometric characteristics of transportation networks. These characteristics are then calculated for 50 cities in the United States and related to their socio-economic properties and travel patterns.

\body

Transportation systems have geometric properties. While their topologic characteristics can be examined as graphs (1–4), complex analysis approaches and more specifically network topological methods (5–10) have recently been used extensively for that purpose (11, 12). Many researchers have focused on presenting a broader picture of transportation networks by showing that they possess general properties such as self-organization (13–17), fractal (18–20), scale-free or power-law distribution (21–25), Zipf's rank law (26–29), or other properties (30–35) to name a few. There are a number of studies of urban systems that have used simulated grid networks for different purposes (13–15, 36, 37). This work focuses on measuring inherent geometric characteristics of urban road networks through studying their grid equivalents, and it is further extended by studying the relationships between the results and their corresponding urban systems' socio-economic characteristics and travel patterns. We first develop the methodology to perform those measurements and then apply it to 50 urban areas in the United States to extract and analyze the characteristics of their road systems.

Cities are complex systems, consisting of a variety of interacting elements. From the time of its inception, an urban settlement goes through an evolutionary process that affects all of its constituents, among them its transportation system. Since a road network grows, expands, and evolves along with and similar to its encompassing urban system, it offers a proper means to study the complexity of its corresponding urban system and to express it using meaningful indicators (38). As Samaniego and Moses have described it: "understanding the topology of urban networks that connect people and places leads to insights into how cities are organized" (39). Moreover, similar to other emerging (14) and self-organizing systems (40), the evolution of road networks is not a simple "product of conscious design" (15), but rather a complex and dynamic process (41) that is the result of the interaction of many different factors. Such influencing parameters include not only the system users and its infrastructure (40), but also topological, morphological, technical, economic, social, and political factors (41), all of which are also determinants of the changes in the road network's encompassing urban system. In fact, even for cities that 'look' different, their transportation systems can demonstrate a variety of similarities (10, 21, 42–44). Based on the above argument, this study contributes to a better understanding of the complex nature of urban road networks by offering a robust and efficient approach that serves as a compliment to other existing methods.

At the first glance, and from a network perspective, a road system is simply seen as a collection of connected segments or links. Understandably, this perspective shifts the main attention towards studying its links as a way of understanding the whole network. This 'link' aspect of urban transportation systems is paramount in terms of geometry and perhaps more closely related to the concept of 'lines' (although not related to Space Syntax (45)). We will therefore look for a *line* indicator that can represent the links in a road system.

An urban road network, however, is more than the sum of its links or lines. Similar to the circulatory system that serves the whole body, a road system serves its encompassing urban area by dividing it into smaller blocks that make it easier to reach every corner of the system. The coverage area of the road network is therefore another important factor to be studied. Thus, we will also represent the coverage area of a given road system by an *area* indicator.

Moreover, the locations where the road segments cross, i.e. their intersections, also play an important role in the daily operation of a road network. For that, their representation should also be a part of any complete study of the complexity of their corresponding urban system. And that provides another objective for this study, which is to find a *point* indicator for a given road network.

Based on the above arguments, an analysis of a road network, as a representative of the complexity of its encompassing urban system, requires three different yet related geometric indicators: *area*, *line*, and *point*. From a mathematical perspective, these three indicators also represent the three main geometric dimensions of an urban system, $D^2$, $D^1$, and $D^0$, respectively. This study offers a unified and systematic approach for the characterization of urban road networks through their *area*, *line*, and *point* indicators, later referred to as *thresholds*.

**Methodology**

In order to explain the methodology towards the development of the three geometric indicators of a given road network, Chicago's urban system is used as an example, for which the process can be summarized in the following three steps. Further details and information are provided in the Supplementary Materials.

Step 1: As the first task, the extent of the urban system for the given city is determined. In the U.S., the commonly-used representation of such an influence area is the city's Metropolitan Statistical Area (MSA)[2]. The choice of MSA not only provides a consistent means for the selection of the extents of an urban area, but it also makes data collection easier as the MSA boundaries are readily available in shapefile format. Figure 1a exhibits Chicago MSA and its road network.

---

[2] MSA is defined as the "geographical region with a relatively high population density at its core and close economic ties throughout the area" (46). This means that "a typical metropolitan area is centered on a single large city that wields substantial influence over the region" (46), e.g. Chicago. More precisely, "Metropolitan Statistical Areas have at least one urbanized area of 50,000 or more population, plus adjacent territory that has a high degree of social and economic integration with the core as measured by commuting ties." (47)

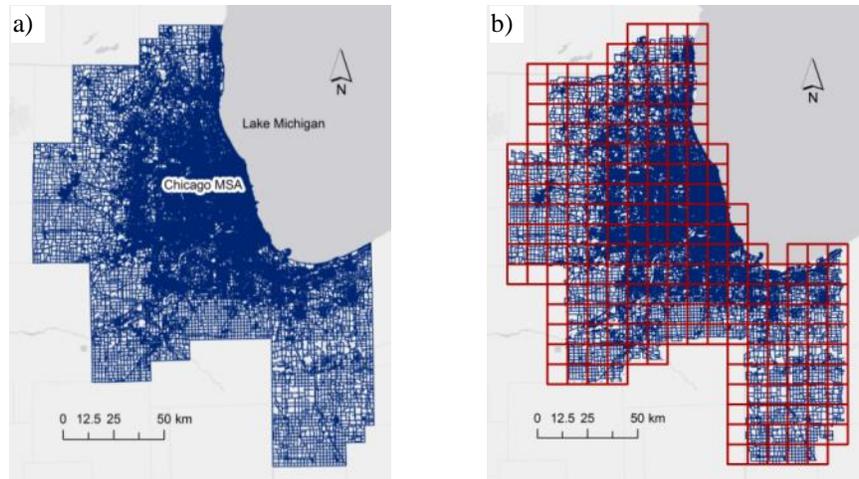

Figure 1. a) Chicago MSA road network, b) Road polygons
and 10x10 km grid network

Instead of the entire MSA area, however, polygons are created wherever the road network within that area creates a closed loop. The rationale for that is to exclude large peripheral areas without roads (see the Supplementary Materials for further information, where we highlight the example of Las Vegas, NV). In other words, the combination of these polygons represents the area serviced by the road segments within them, for which the total area is calculated. Moreover, for the given road network, the total road length and total number of intersections are also calculated.

Step 2: This step involves successive creation of grids with varying cell sizes, overlaying them on the road network, and extracting the cells needed to cover all the road segments. Then, for each grid the total coverage area, total length of links, and total number of nodes are calculated. Figure 1b demonstrates the creation of such a grid with 10 km x 10 km cells that covers Chicago MSA road network. Note that this process resembles the box-counting methodology in fractal analysis, although here different information is collected from the results.

Step 3: The final task involves a comparison of the values obtained from Steps 1 and 2. The goal is to find the specific grids that are equivalent to the given road network with respect to total coverage area, total road length, and total number of intersections. As mentioned before, those criteria represent the given road network's area, line, and point characteristics, successively. The idea is that while a given urban road system might have an irregular configuration, something which is a part of its complex identity, one should be able to find equivalent grid networks that possesses the same area, line, or point geometric characteristics.

Naturally, there are more than one equivalent grid network that satisfy the condition for any of the above geometric indicators. An additional condition must therefore be set to result in a unique grid network. For that, we require the coverage area of the grid network to cover all the road segments of the urban road network under study, which is essentially already achieved in Step 2. The block size of the equivalent grid network will then be considered as the indicator, or as called hereafter: the "*threshold*", for its corresponding geometric characteristic (area or line or point).

The procedure explained above is applied to the Chicago MSA road network. Due to its dense configuration, however, only a south-western section of the road network along with its equivalent grid networks are magnified and demonstrated in Figure 2.

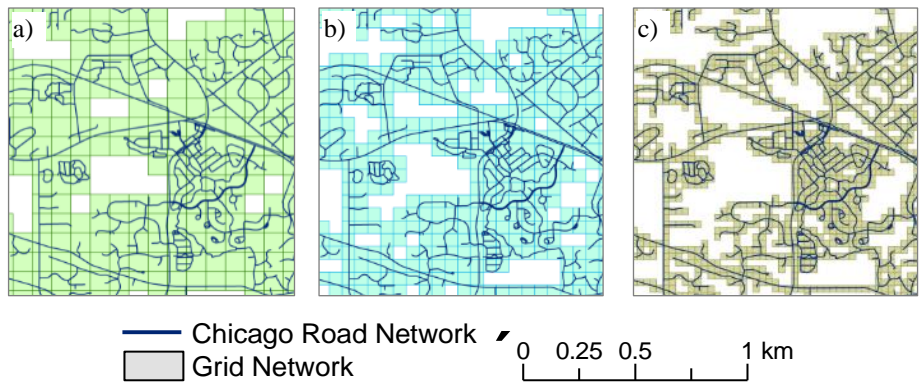

Figure 2. Comparison of Chicago road network and different grids with a) equal area, b) equal road length, and c) equal number of intersections. To explain their differences, the process involves overlaying grids with gradually decreasing block sizes over the original road system. At the beginning, the area covered by the grid is larger than the corresponding area of the road network under study. As the cell size of the grid is gradually reduced, at some point the two areas become equal (Figure 2a). At that very moment, the grid network crosses a threshold. Since it marks the point where the two networks are equivalent in an "area" dimensional perspective, the grid network's block size is then designated as the *area threshold*. After that point, the focus shifts to the comparison of the total road lengths of the two networks. As the grid network's block size becomes smaller and smaller, its total road length gradually increases, up to a point at which it becomes equal to the total road length of the original network (Figure 2b). That moment marks another threshold, at which the block size of the grid network is designated as the *line threshold*, i.e. when the two networks are equivalent in a "line" dimensional sense. The same process continues further, until a point when the total numbers of intersections (points) in both networks become equal (Figure 2c). That marks the third threshold, at which the block size of the grid network is designated as the *point threshold*. At that very moment, the two networks are equivalent in a "point" dimensional sense.

As discussed before, a given urban road network can be examined from different perspectives. One is the area it encompasses or serves. Another one is the links (lines) that facilitate the services it provides. And the third one is the intersections (points) that in turn facilitate the transfer of services between links (lines). Measuring these three components, and their corresponding thresholds (as explained above), can help better understand the characteristics of the urban system itself.

In order to find the three area, line, and point thresholds accurately, the following approach is taken.

For a given urban road network, its coverage area ($A$), total road length ($L$), and total number of intersections ($P$) can be calculated and extracted from its shapefile, easily obtainable from Census TIGER/Lines dataset (48).

In comparison, for any chosen grid network with a block size of $\varepsilon$, the area it serves ($a$), the total road length it consists of ($l$), and the total number of intersections that it has ($p$), can also be extracted from its shapefile.

Instead of comparing the two sets of numbers, the grid network values are normalized by dividing them by the road network's corresponding values and are then compared with the unity (one), i.e. plots of $a/A$, $l/L$, and $p/P$, are drawn and intersected with a horizontal line with the value of 1. At the point of intersection, the block size ($\varepsilon$) of the grid network is extracted and reported as the corresponding threshold. Examples of the diagrams for the area, line, and point thresholds for the Chicago MSA road network are presented in Figure 3.

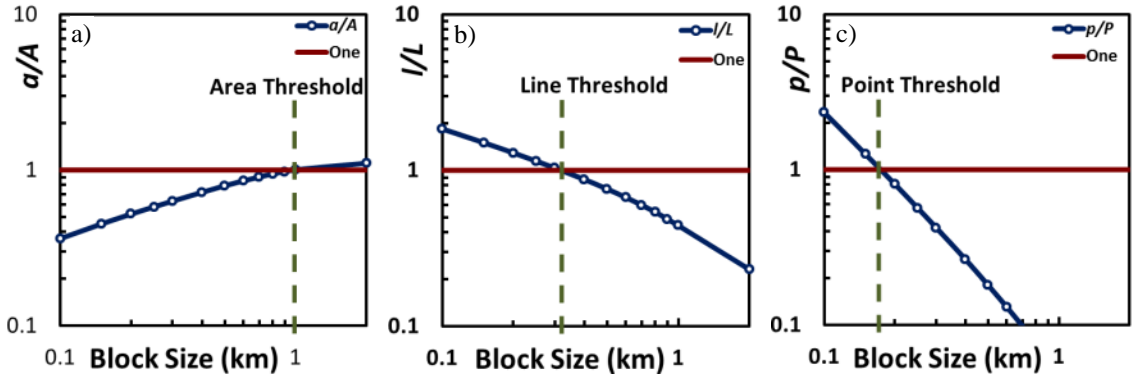

Figure 3. Determining the three thresholds for Chicago MSA road network:

a) Area, b) Line, and c) Point Thresholds

**Results and Discussion**

Similar steps were executed for a total of 50 urban areas across the U.S. (see the Supplementary Materials for a complete list of cities as well as individual results). These cities cover a wide and diverse range of parameters such as road network structure, topology, morphology, history, size, population, area, and socio-economic conditions. The results of the analyses performed are presented in Figure 4 in the form of three maps, showing the geospatial variations of the three thresholds (area, line, and point) calculated for those urban areas.

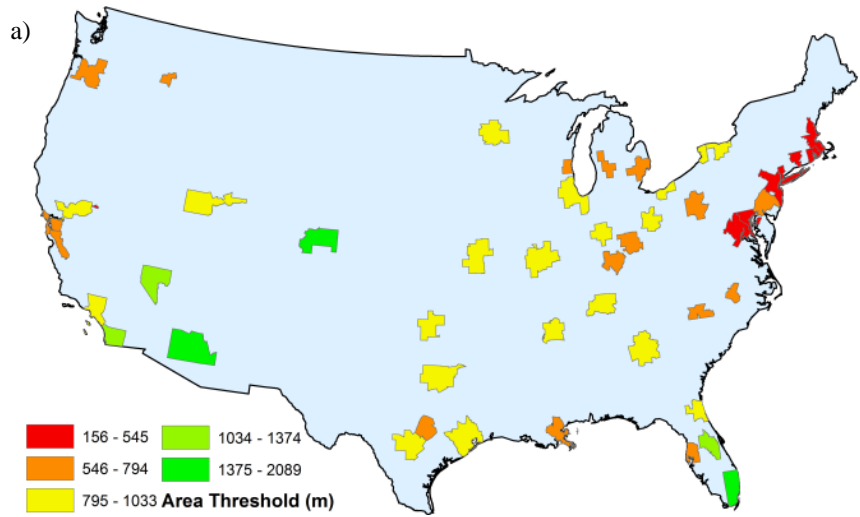

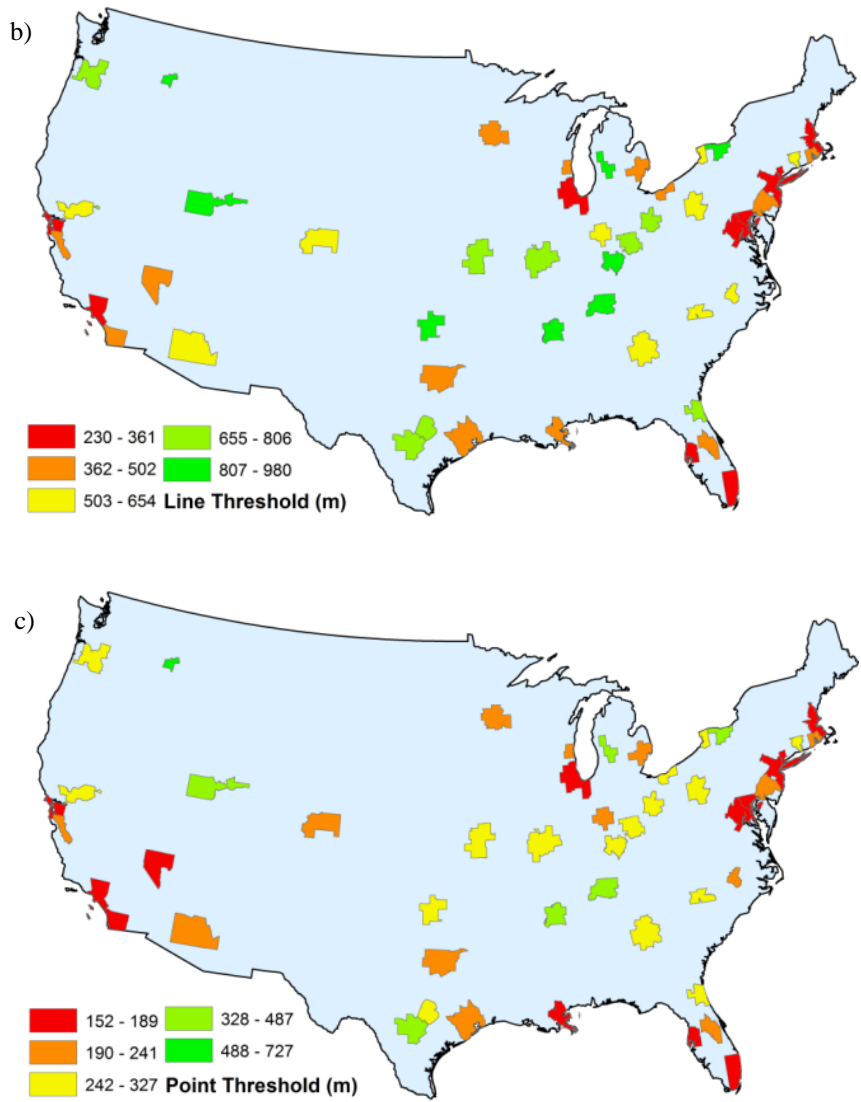

Figure 4. a) Area, b) Line, and c) Point thresholds. Area threshold: this figure shows lower values for older cities (mostly in the north-eastern states) as compared to those for younger cities. This difference is related to the advent of the motorized transportation in the 20$^{th}$ century. 'Older' cities tend to be more walkable and have smaller blocks, while 'younger' cities tend to have larger block sizes. A comparison between Phoenix, AZ with Chicago, IL that have the largest and medium area thresholds, respectively, sheds light on this fact (please refer to the Supplementary Materials). Line threshold: different from the previous figure, we see that the line thresholds for the cities along the costal line are smaller than for the cities inside the country. The reason is partly due to the fact that coastal
Page **9** of **42**

cities often perform as logistical hubs (e.g., ports) and thus are centers of import and export activities. As a result, their road networks are more compact and have more uniform road segments as compared to inland cities that have larger variations in their road segment lengths. A comparison of the length variations within the road networks of Salt Lake City, UT with Chicago, IL that have the largest and medium line thresholds, respectively, presents a visual explanation of this characteristic (please refer to the Supplementary Materials). Point threshold: while we might expect to see the same trend for the point threshold as the line threshold, since intersections are merely where the roads intersect, this is not always the case. A good example is Denver, Colorado, that has a mid-range line threshold, but a small point threshold. One of the factors affecting the point threshold is the way the intersections are created, i.e. 6- or 4- way intersections as compared to T- intersections or cul-du-sacs, each affecting the point-threshold differently. This means that cities with similar line thresholds could have different point threshold, and vice versa.

The above figures demonstrate interesting and insightful aspects of the diversity of the inner complexity of the urban systems studied here. Of relevance, overall no single indicator can completely capture and describe the complexities at play. This emphasizes the fact that any given urban system has its own unique multi-dimensional complex characteristics, all of which are needed to gain a complete picture of its urban characteristics.

In order to better present and visually compare the thresholds calculated for the cities studied in this work, all the values obtained are plotted in one diagram (Figure 5). The figure clearly shows that each threshold has its own variation and no two thresholds are behaving similarly, again a manifestation of the complex nature of urban systems and their road networks.

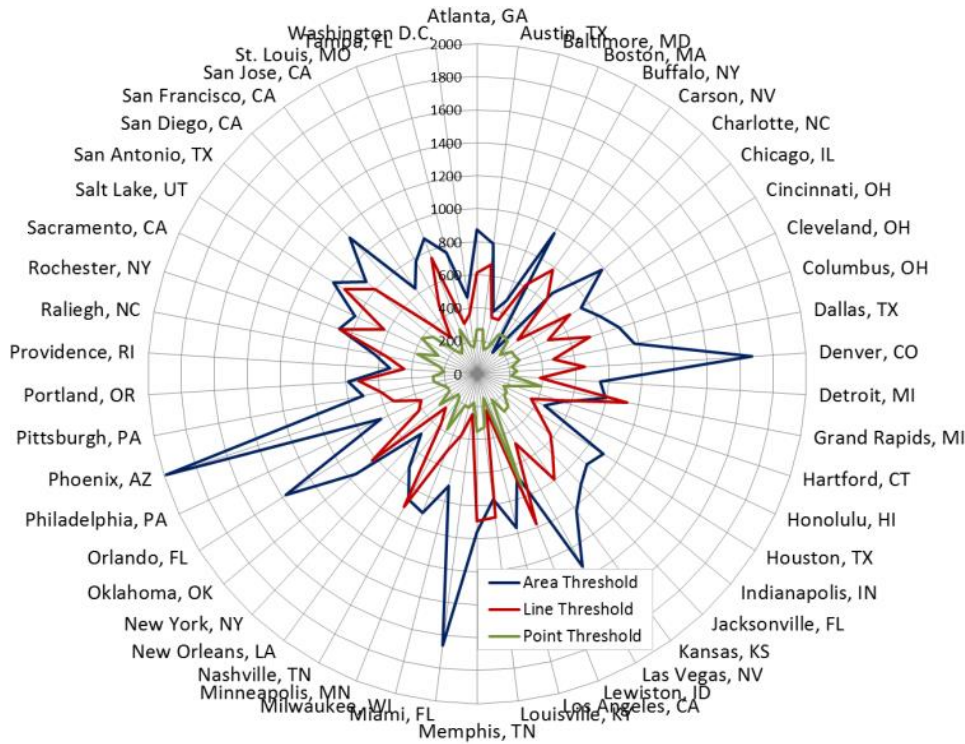

Figure 5. Area, line, and point thresholds for 50 U.S. urban areas.

The significance of the three thresholds found in this work were further investigated through analyzing their relationships with several socio-economic parameters as well as travel patterns related to their corresponding urban areas are studied here.

A plot of the area threshold versus age of the urban systems (49) studied here is presented in Figure 6.

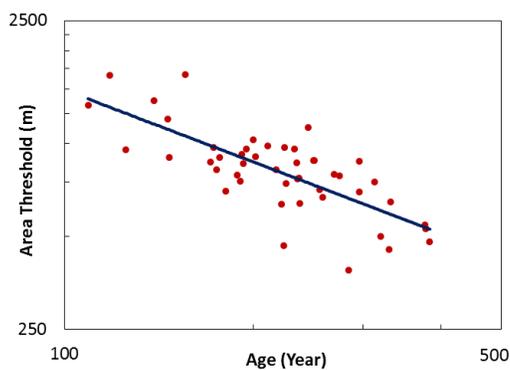

Figure 6. Relationship between the area threshold and the age of the urban system. The figure shows a power law trend (*Area Threshold* = 52057 $Age^{-0.772}$, $R^2$ = 0.51, and |t-score| = 6.8), which means that the older a city is, the shorter its area threshold will be. This supports the fact that in older cities the polygons

Page **11** of **42**

created by road networks are smaller due to their more developed state, while in younger cities one would see larger polygon sizes. This figure is able to capture nearly two hundred years of urban and regional planning theory and the advent of motorized transportation as discussed earlier.

From another perspective we witness a relationship between population density and line and point thresholds, as shown in Figure 7. This phenomenon is common and expected (52–54), since, if other conditions remain the same, neighborhoods with smaller blocks (i.e. higher road and intersection density, as compared to larger blocks) tend to create safer environments and thus attract more people, hence higher population density.

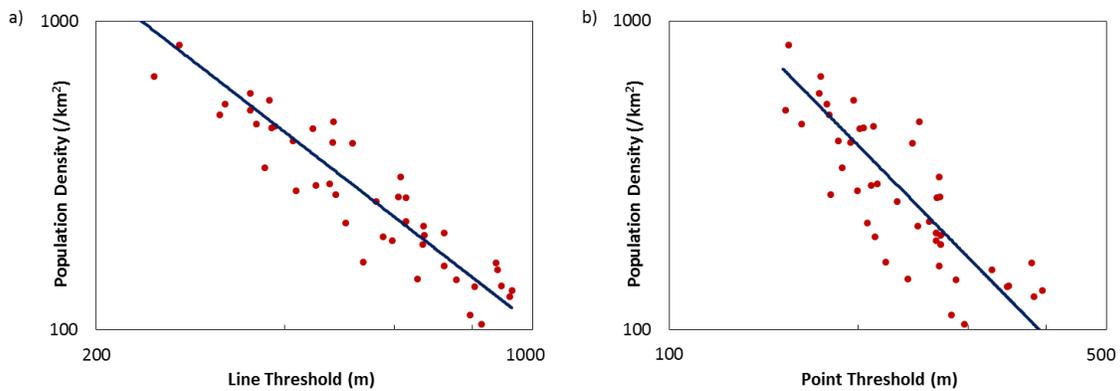

Figure 7. Relationship between Population density and a) Line and b) Point thresholds. Figures 7a shows that an increase in line threshold, which means larger block size, has a negative power law impact on population density (*Population density* = $5 \times 10^6$ *Line threshold*$^{-1.566}$, $R^2 = 0.56$, and |t-score| = 7.6). The reason is that longer road segments, i.e., larger block sizes, essentially translate into larger residential units. Similarly, population density is affected by point threshold (*Population Density* = $2 \times 10^7$ *Point threshold*$^{-2.061}$, $R^2 = 0.43$, and |t-score| = 5.8), as shown in Figure 7b. This shows the fact that closer and denser intersections translate into city blocks that are smaller and thus more suitable for housing with higher concentration of people per area.

Using the 2010 American Community Survey (ACS) data (50), we find that many travel patterns within the U.S. have power law relationships with the line threshold. Figure 8 exhibits the variations of the average travel time for all modes and also total transit travel time with respect to the line threshold. Other travel patterns found to possess similar trends, including all-modes total travel time, total number of trips, and total number of transit trips.

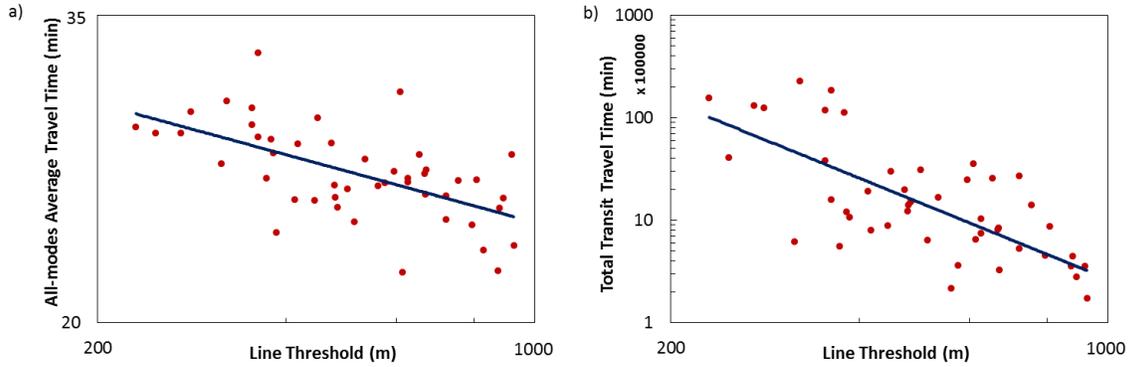

Figure 8. Relationship between the a) Average travel time for all modes, and b) Total transit travel time, and the Line threshold. The power law trend seen in Figure 8a (*All-modes avg. travel time* = 60.725 *Line threshold*$^{-0.134}$, $R^2$ = 0.38, and |t-score| = 5.2) shows that as the line threshold increases, the average travel time for all modes decreases. The reason is that an increase in the length of the road segments, which partially represents the existence of freeways and thus lower road density, results in a higher car use as the dominant choice of transportation mode in the U.S. A similar trend exists for the reduction in the use of public transit, shown in Figure 8b (*Total transit travel time* = 7x10$^{12}$ *Line threshold*$^{-2.475}$, $R^2$ = 0.30, and |t-score| = 4.4). In this instance, we use total as opposed to average travel time since cities with denser road networks tend to generate more as well as longer transit trips. As a result the total number of transit trips and thus the total transit travel time drop as line threshold increases.

As for the point threshold, studies (51, 52) have shown that denser road networks, which translate into closer and more compact intersections, support active modes of transportation, including walking. Walking data from 2010 American Community Survey (ACS) supports this idea, as shown in Figure 9.

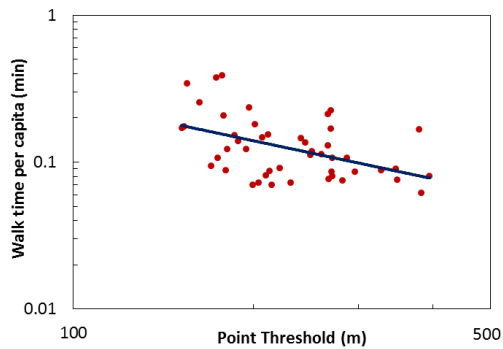

Figure 9. Relationship between Walk time per capita and Point threshold. Based on this figure, urban areas with shorter point thresholds, i.e. with more and closer intersections, have higher walk time per capita (*Walk time per capita* = 13.405 *Point threshold*$^{-0.861}$, $R^2$ = 0.38, and |t-score| = 3.1). This result simply reflects that among other parameters the closer the intersections, the more encouraging and supportive the environment is for pedestrians. In other words, pedestrians are willing to walk longer distances.

**Conclusion**

This study offered a complementary perspective into the complex nature of urban systems via geometric properties of their road networks. By creating grid networks of varying block sizes and overlaying them on the road networks under study, three indicators were extracted, each representing an individual geometric property of the network. Together, the area, line, and point thresholds obtained through the method developed in this study succeed in capturing important and complex characteristics of an urban system. While two cities may share similarities for one of the thresholds, they may not be similar with respect to the other two, thus allowing us to capture their unique properties from different perspectives. In this study, we first developed a methodology to measure these three thresholds, which we then applied to 50 U.S. urban systems with a wide variety of characteristics. We also showed that there are correlations between the thresholds defined and extracted in this study and the socio-economic characteristics as well as travel patterns for a given urban area.

**Acknowledgments**

The authors would like to thank Mrs. Elham Peiravian for extensive help with data entry and data preparation. This research was partly funded by the Department of Civil and Materials Engineering at the University of Illinois at Chicago.

# SUPPLEMENTARY INFORMATION

A. List of 50 U.S. urban systems studied
B. Rational for choosing road polygons over MSA polygon (Case study: Las Vegas, NV)
C. Explanation for Step 2 of the Methodology (Case study: Chicago, IL)
D. Road networks for 50 U.S. urban systems
E. Discussion of area threshold
F. Discussion of line threshold
G. Area, point, and line thresholds for 50 U.S. road systems

A. List of 50 U.S. urban systems studied

| Urban Area, State | Founded in[1] | Population[2] | Area (km²)[3] | Pop Density | Road Length (km)[3] | # of Intersections[3] |
|---|---|---|---|---|---|---|
| Atlanta, GA | 1843 | 5486738 | 20306.8 | 270.2 | 67215.1 | 243462 |
| Austin, TX | 1835 | 1784094 | 9440.8 | 189.0 | 30382.0 | 111234 |
| Baltimore, MD | 1729 | 2895944 | 5624.9 | 514.8 | 35556.3 | 220784 |
| Boston, MA | 1630 | 4892136 | 8368.7 | 584.6 | 49139.9 | 261949 |
| Buffalo, NY | 1789 | 1191744 | 3821.4 | 311.9 | 12293.0 | 41429 |
| Carson, NV | 1858 | 87743 | 109.0 | 804.9 | 900.6 | 3045 |
| Charlotte, NC | 1755 | 1927130 | 7177.5 | 268.5 | 24978.8 | 93988 |
| Chicago, IL | 1803 | 9594379 | 17783.6 | 539.5 | 86788.9 | 396704 |
| Cincinnati, OH | 1788 | 2252951 | 10398.8 | 216.7 | 33834.5 | 141744 |
| Cleveland, OH | 1796 | 2272776 | 4827.5 | 470.8 | 19472.2 | 64630 |
| Columbus, OH | 1812 | 1949603 | 9483.2 | 205.6 | 27764.3 | 106156 |
| Dallas, TX | 1841 | 6501589 | 21833.1 | 297.8 | 83815.2 | 350762 |
| Denver, CO | 1858 | 2666592 | 18262.0 | 146.0 | 46547.0 | 182157 |
| Detroit, MI | 1701 | 4369224 | 9664.6 | 452.1 | 46880.4 | 187960 |
| Grand Rapids, MI | 1825 | 895227 | 6665.8 | 134.3 | 16684.6 | 42990 |
| Hartford, CT | 1637 | 1400709 | 3487.6 | 401.6 | 14992.7 | 56695 |
| Honolulu, HI | 1809 | 953207 | 775.4 | 1229.3 | 4678.9 | 22904 |
| Houston, TX | 1837 | 6052475 | 20585.7 | 294.0 | 83365.0 | 353831 |
| Indianapolis, IN | 1821 | 1856996 | 9289.1 | 199.9 | 32389.9 | 150469 |
| Jacksonville, FL | 1822 | 1451740 | 7182.3 | 202.1 | 22067.4 | 76396 |
| Kansas City, KS | 1868 | 2138010 | 19148.1 | 111.7 | 50639.6 | 184748 |
| Las Vegas, NV | 1905 | 2010951 | 7330.1 | 274.3 | 20926.8 | 104925 |
| Lewiston, ID | 1861 | 85096 | 2104.6 | 40.4 | 4206.1 | 6334 |
| Los Angeles, CA | 1781 | 13059105 | 10913.2 | 1196.6 | 70096.7 | 335638 |
| Louisville, KY | 1778 | 1443801 | 9227.8 | 156.5 | 24453.7 | 82680 |
| Memphis, TN | 1819 | 1398172 | 10049.2 | 139.1 | 25028.4 | 74462 |
| Miami, FL | 1896 | 5571523 | 8410.3 | 662.5 | 42827.1 | 178680 |

| Urban Area, State | Founded in[1] | Population[2] | Area (km$^2$)[3] | Pop Density | Road Length (km)[3] | # of Intersections[3] |
|---|---|---|---|---|---|---|
| Milwaukee, WI | 1833 | 1602022 | 3507.8 | 456.7 | 17207.1 | 66802 |
| Minneapolis, MN | 1867 | 3412291 | 15365.8 | 222.1 | 57532.0 | 259788 |
| Nashville, TN | 1779 | 1740134 | 13588.3 | 128.1 | 32653.8 | 90700 |
| New Orleans, LA | 1718 | 1247062 | 3715.5 | 335.6 | 18340.7 | 83361 |
| New York, NY | 1624 | 19217139 | 15551.5 | 1235.7 | 105344.0 | 499969 |
| Oklahoma, OK | 1889 | 1359027 | 13051.0 | 104.1 | 34167.6 | 120303 |
| Orlando, FL | 1875 | 2257901 | 7996.6 | 282.4 | 28876.5 | 123076 |
| Philadelphia, PA | 1682 | 6234336 | 11271.7 | 553.1 | 58104.3 | 256023 |
| Phoenix, AZ | 1868 | 4262838 | 25763.0 | 165.5 | 60738.6 | 241836 |
| Pittsburgh, PA | 1758 | 2503836 | 12859.9 | 194.7 | 45196.4 | 167027 |
| Portland, OR | 1845 | 2363554 | 14669.4 | 161.1 | 44544.0 | 174765 |
| Providence, RI | 1636 | 1695760 | 3773.5 | 449.4 | 18431.5 | 83871 |
| Raliegh, NC | 1792 | 1258825 | 4830.5 | 260.6 | 18678.0 | 81802 |
| Rochester, NY | 1803 | 1159166 | 7037.2 | 164.7 | 17863.9 | 47275 |
| Sacramento, CA | 1839 | 2277843 | 10167.0 | 224.0 | 34020.6 | 124839 |
| Salt Lake, UT | 1847 | 1246208 | 10895.1 | 114.4 | 22387.0 | 59736 |
| San Antonio, TX | 1718 | 2239307 | 16213.5 | 138.1 | 44137.5 | 127773 |
| San Diego, CA | 1769 | 3144425 | 7668.0 | 410.1 | 29499.1 | 144194 |
| San Francisco, CA | 1776 | 4472992 | 5352.1 | 835.7 | 33483.0 | 172400 |
| San Jose, CA | 1777 | 1992872 | 4921.2 | 405.0 | 19824.6 | 93610 |
| St. Louis, MO | 1763 | 2934412 | 20184.1 | 145.4 | 57670.8 | 205269 |
| Tampa, FL | 1823 | 2858974 | 5756.8 | 496.6 | 31421.2 | 143714 |
| Washington D.C. | 1790 | 5916033 | 12735.0 | 464.5 | 74190.6 | 437470 |

1. Wikipedia, Accessed 2014-06: http://www.wikipedia.org/
2. U.S. Census Bureau American FactFinder, 2010: http://factfinder2.census.gov/
3. Calculated from U.S. Census Bureau TIGER/Line Shapefiles, 2010: https://www.census.gov/geo/maps-data/data/tiger-line.html

B. Rational for choosing road polygons over the MSA polygon (Case study: Las Vegas, NV)

Polygons are created for the area threshold analysis wherever the road network within that area created a closed loop. The rationale for that is to exclude outer road segments that extend beyond the built environment of a MSA that follows county boundaries. This is desirable since full MSA areas easily artificially inflate the area of an urban system instead of focusing on the area serviced by the road network. The figure below demonstrates the difference between the road polygons area created using the above approach versus the MSA area for the Las Vegas MSA, which makes a substantial difference in the service area to be analyzed.

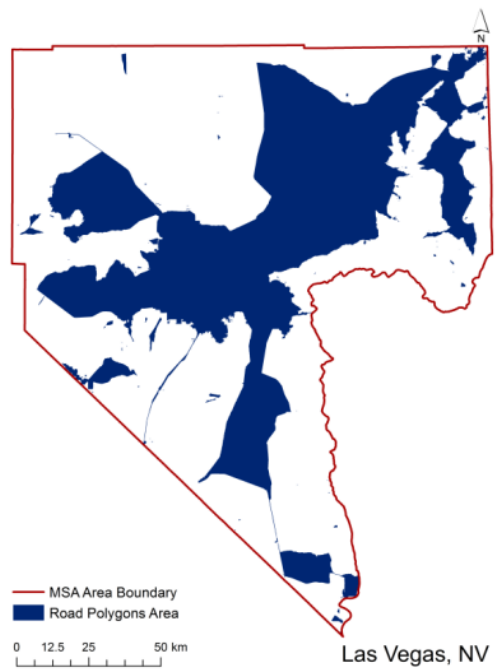

Figure B. Difference between the service area of the road system created by closed road polygons versus the MSA area for the city of Las Vegas.

C. Explanation for Step 2 of the Methodology (Case study: Chicago, IL)

In Step 1 of the methodology, as explained before, at first the MSA of the given urban area was chosen as its extent. In Step 2, successive grids with varying block sizes were created and then overlaid on the road network, from which the cells needed to cover all the road segments within the MSA were extracted. Figures below demonstrate the creation of grids with cells ranging from 10 km to 100 m for Chicago MSA road network. They show how the grid network evolves towards the real road network. During this process, there are thresholds at which the grid network becomes equivalent to the road network from area, line, and point dimensional perspectives. The last figure demonstrates the real road network for visual comparison.

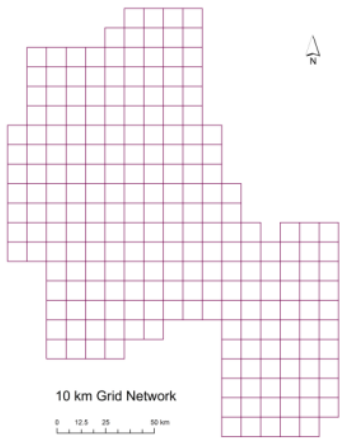 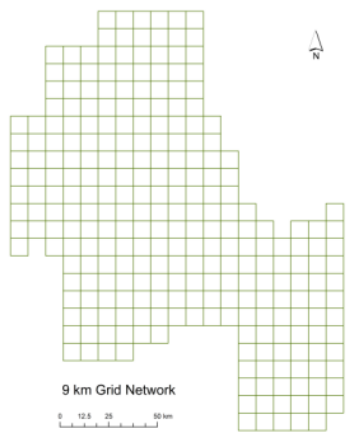 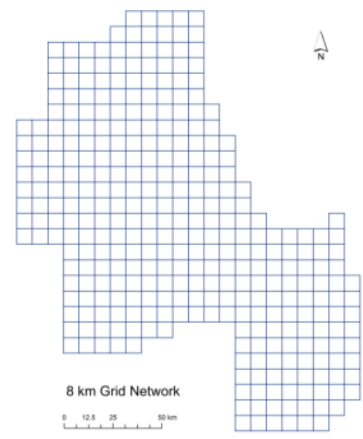

a) 10 km Grid Networkb) 9 km Grid Networkc) 8 km Grid Network

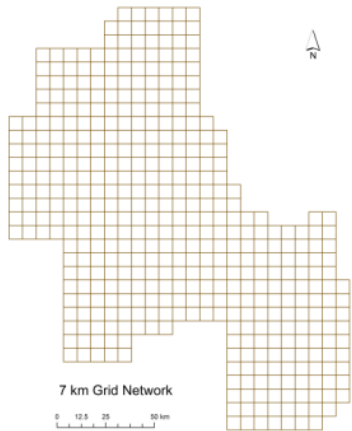 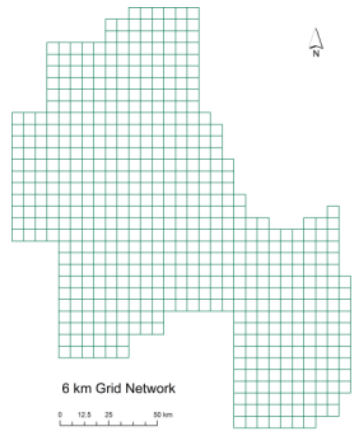 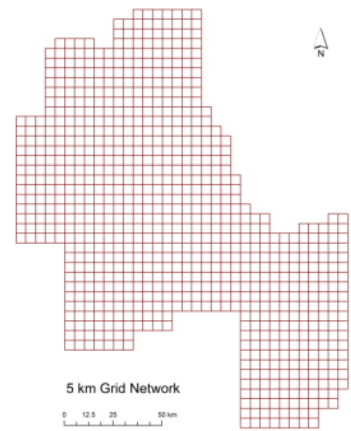

d) 7 km Grid Networke) 6 km Grid Networkf) 5 km Grid Network

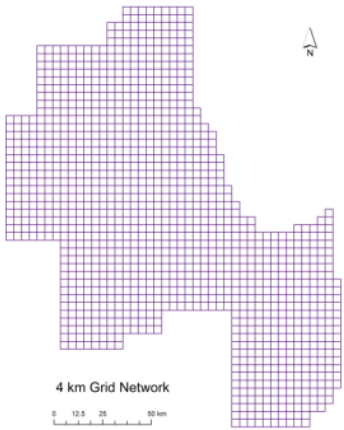 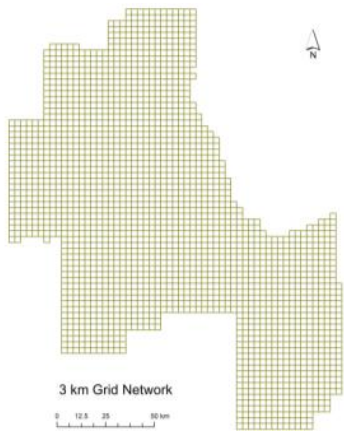 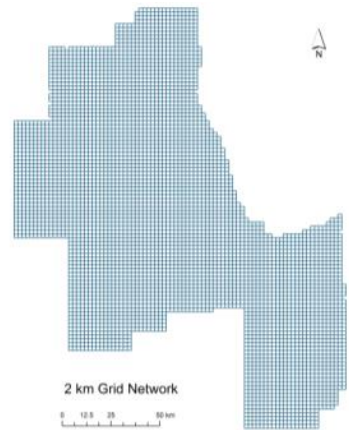

g) 4 km Grid Network    h) 3 km Grid Network    i) 2 km Grid Network

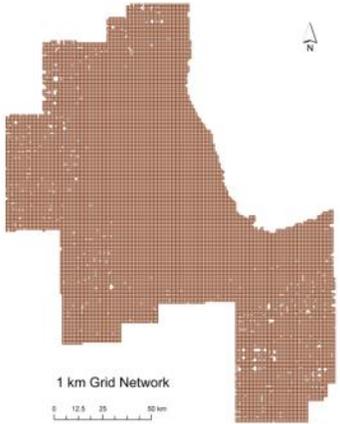 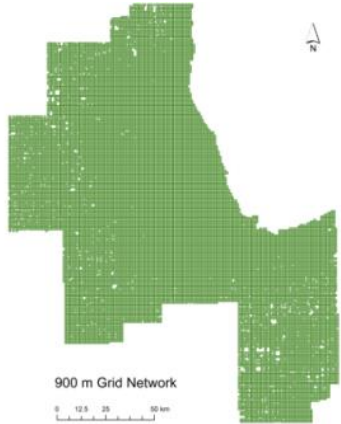 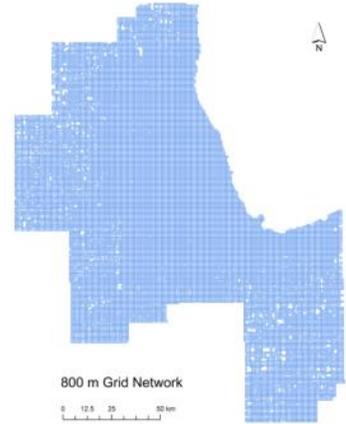

j) 1 km Grid Network    k) 900 m Grid Network    l) 800 m Grid Network

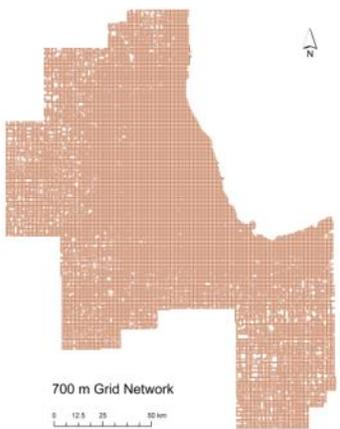 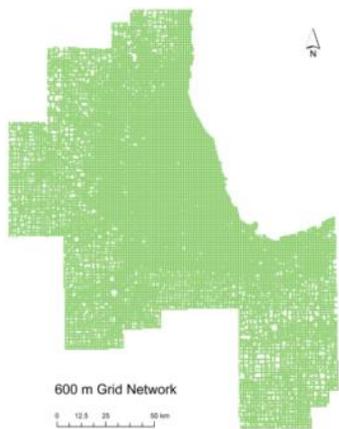 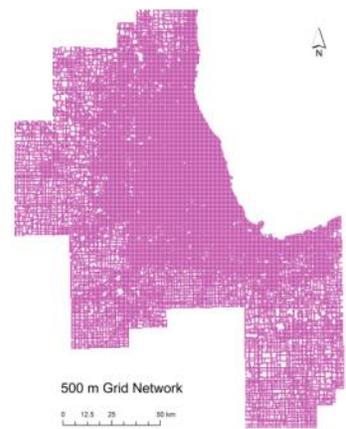

m) 700 m Grid Network    n) 600 m Grid Network    0) 500 m Grid Network

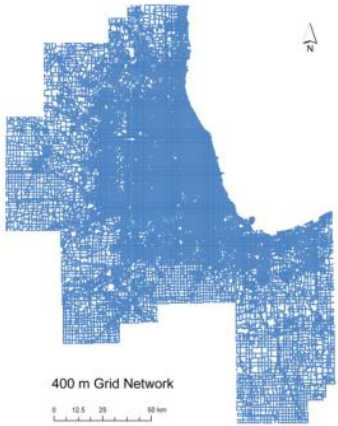
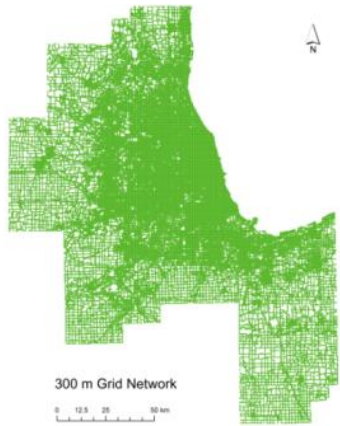
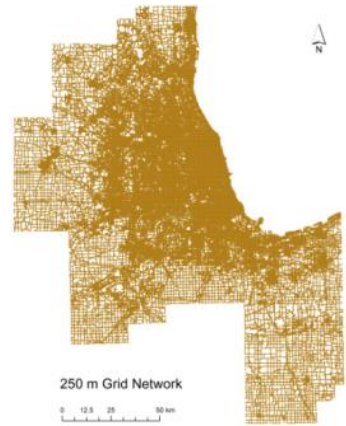

p) 400 m Grid Network     q) 300 m Grid Network     r) 250 m Grid Network

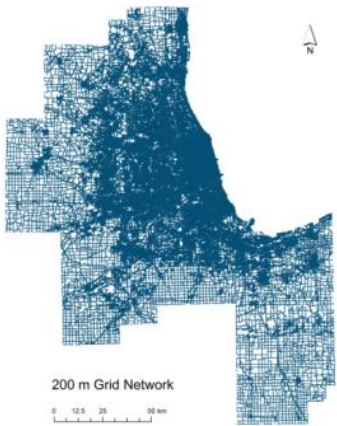
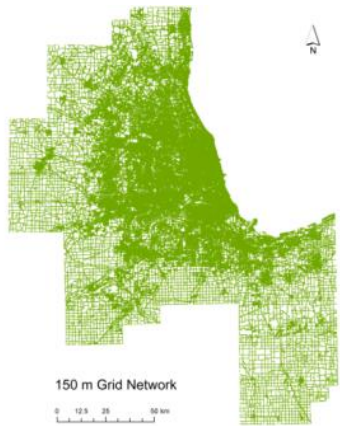
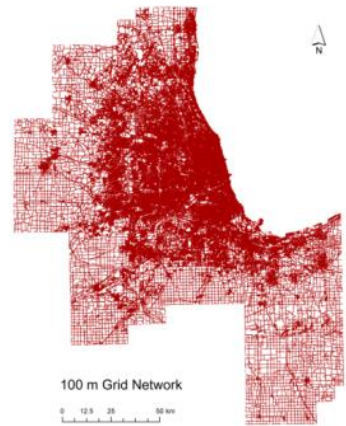

s) 200 m Grid Network     t) 150 m Grid Network     u) 100 m Grid Network

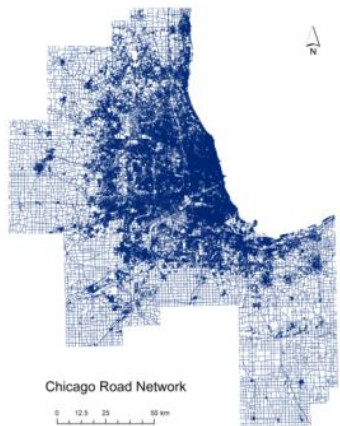

v) Chicago Road Network

Figure C. Evolution of grid networks for Chicago MSA road network

D. Road networks for 50 U.S. urban systems

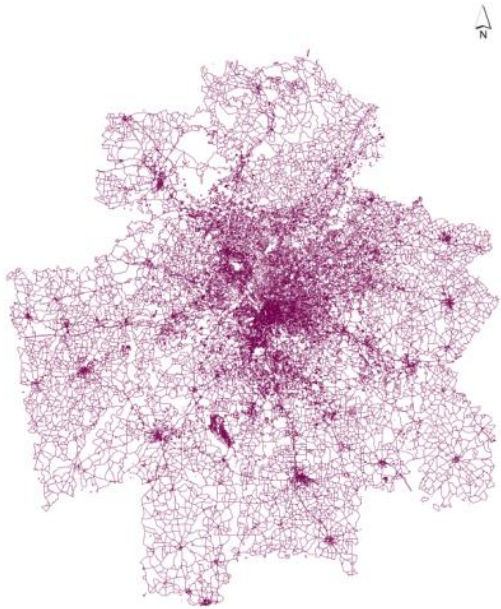
Atlanta, GA

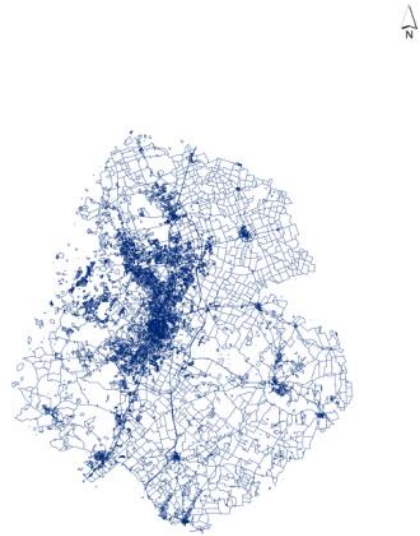
Austin, TX

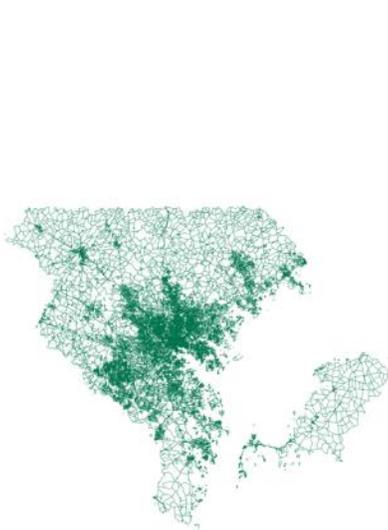
Baltimore, MD

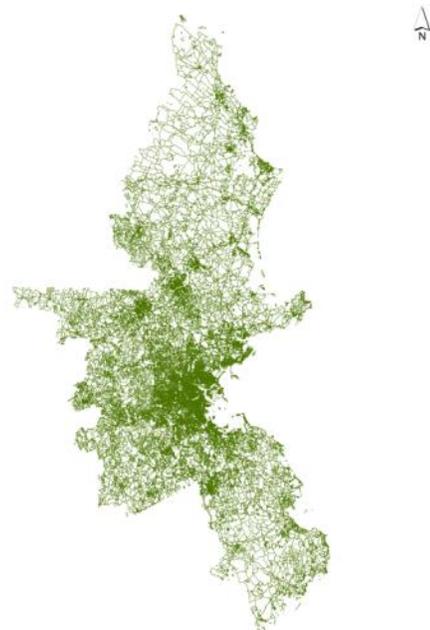
Boston, MA

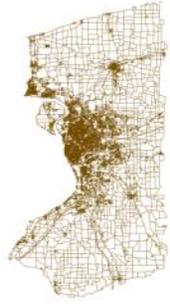

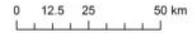
Buffalo, NY
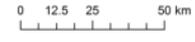
Carson, NV

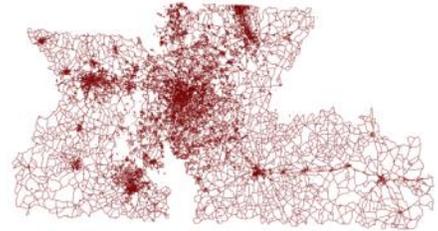
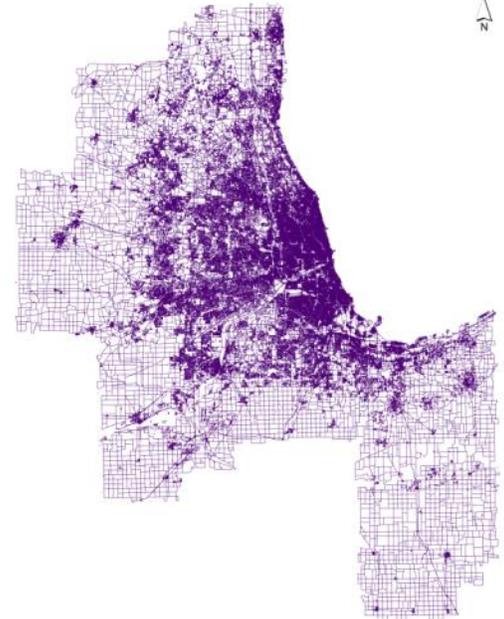
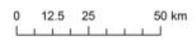
Charlotte, NC
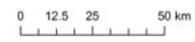
Chicago, IL

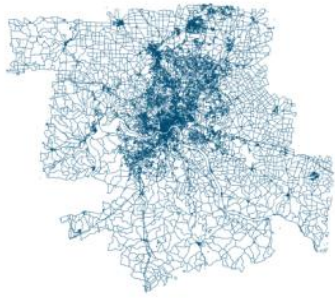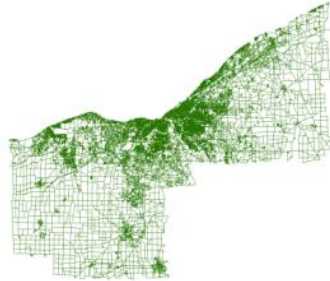

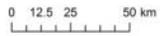 Cincinnati, OH

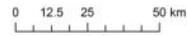 Cleveland, OH

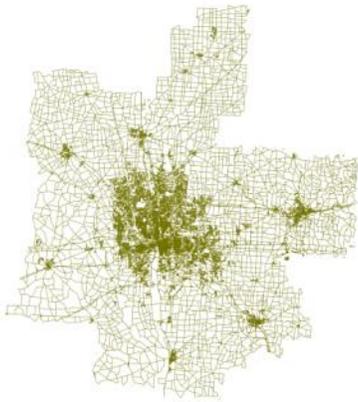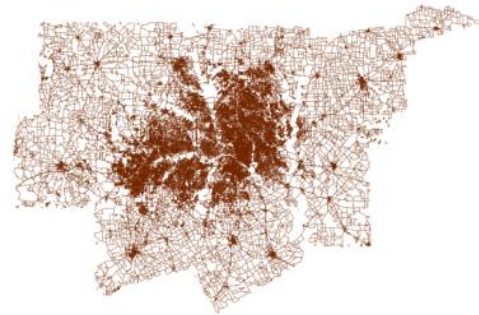

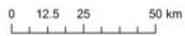 Columbus, OH

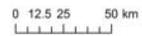 Dallas, TX

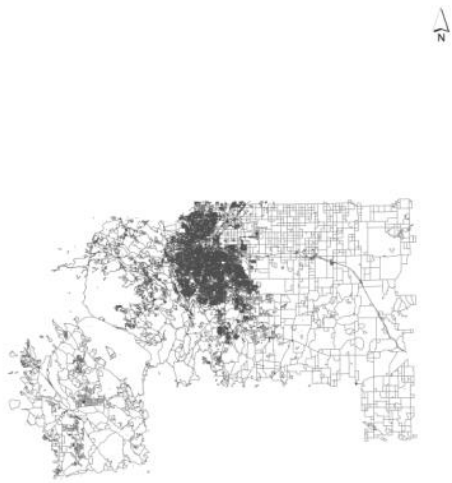
Denver, CO

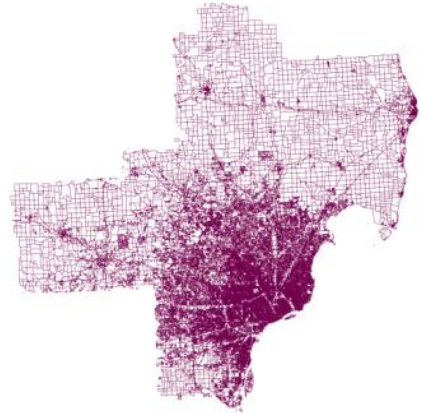
Detroit, MI

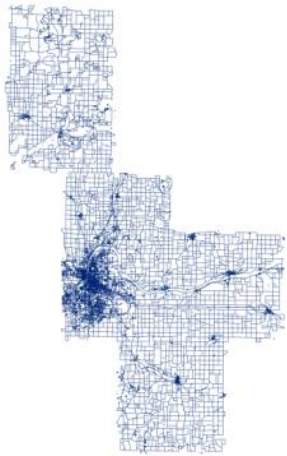
Grand Rapids, MI

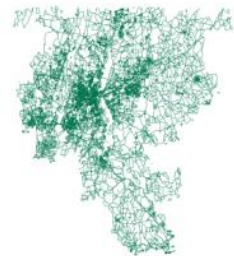
Hartford, CT

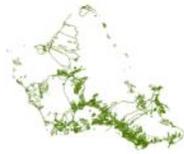
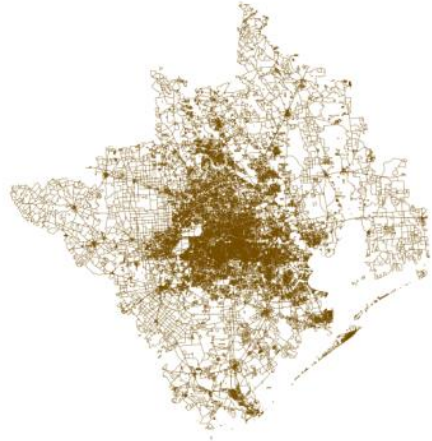
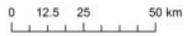
Honolulu, HI
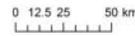
Houston, TX

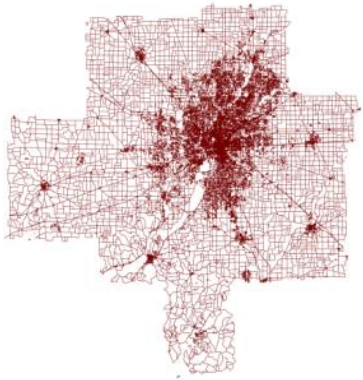
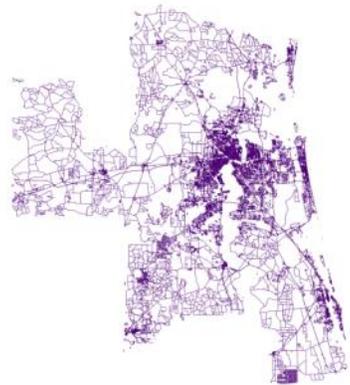
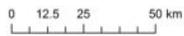
Indianapolis, IN
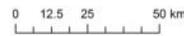
Jacksonville, FL

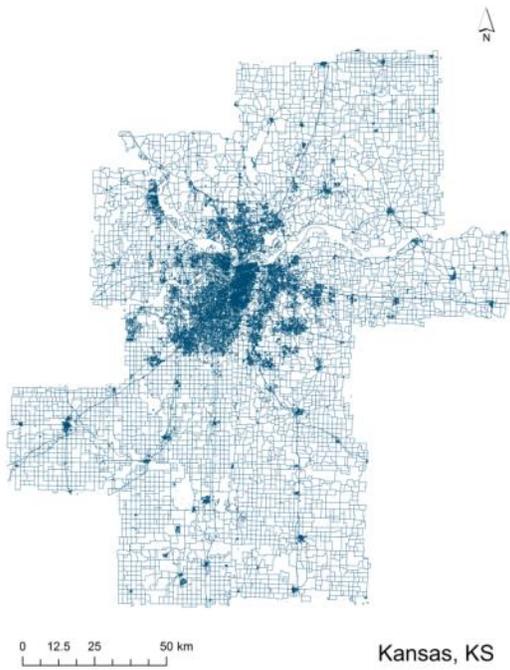
Kansas, KS
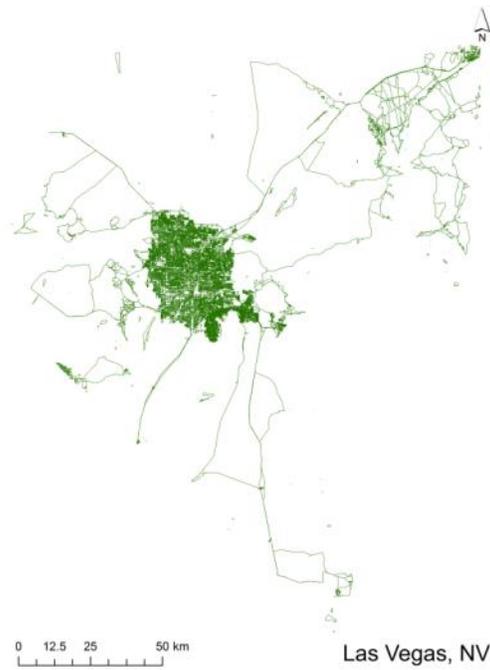
Las Vegas, NV
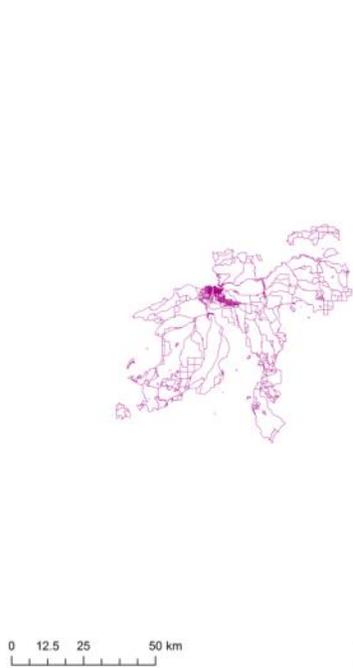
Lewiston, ID
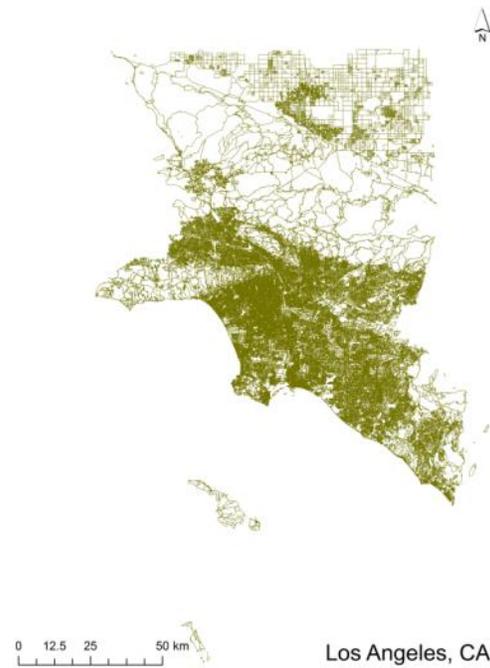
Los Angeles, CA

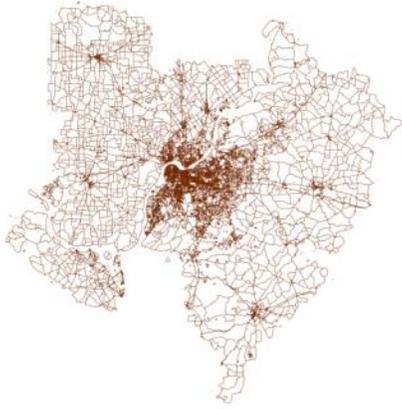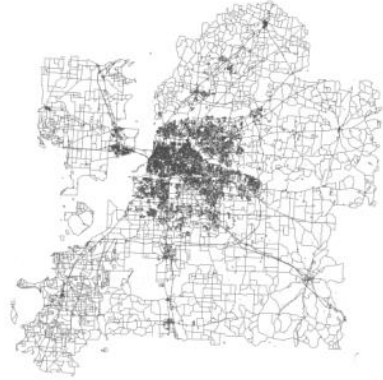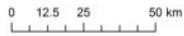  Louisville, KY
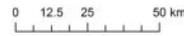  Memphis, TN

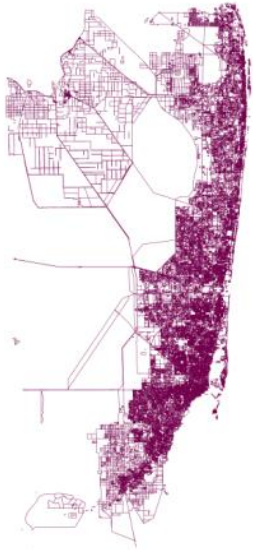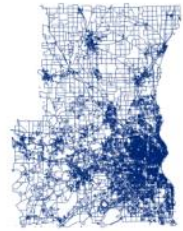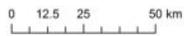  Miami, FL
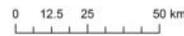  Milwaukee, WI

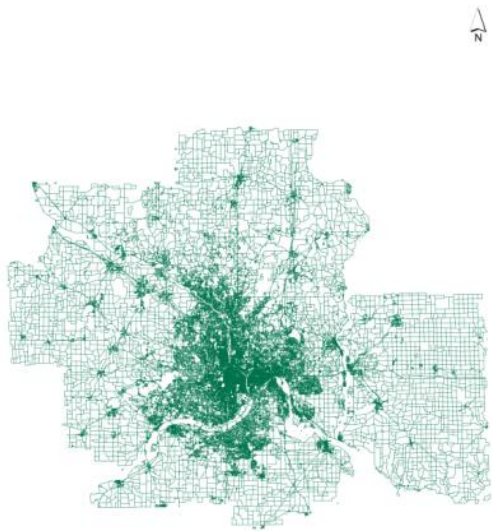
Minneapolis, MN

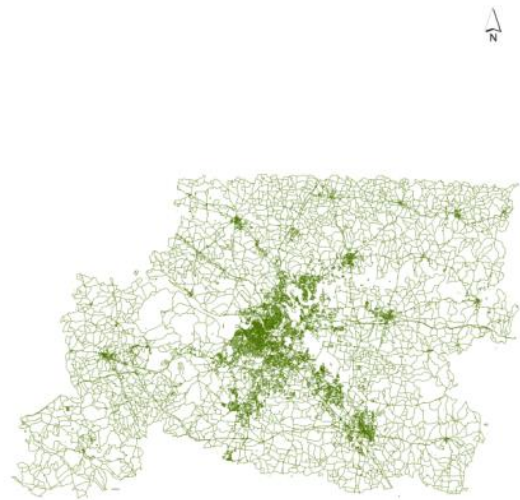
Nashville, TN

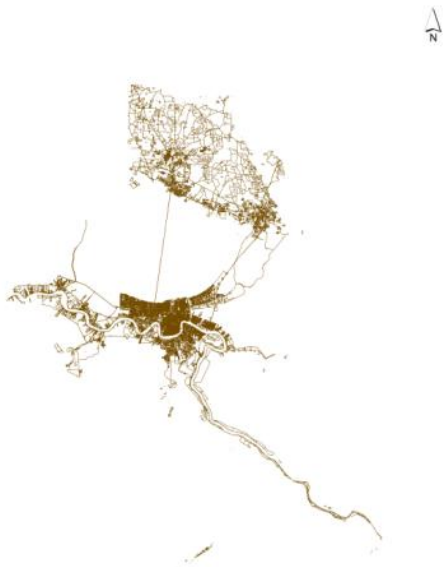
New Orleans, LA

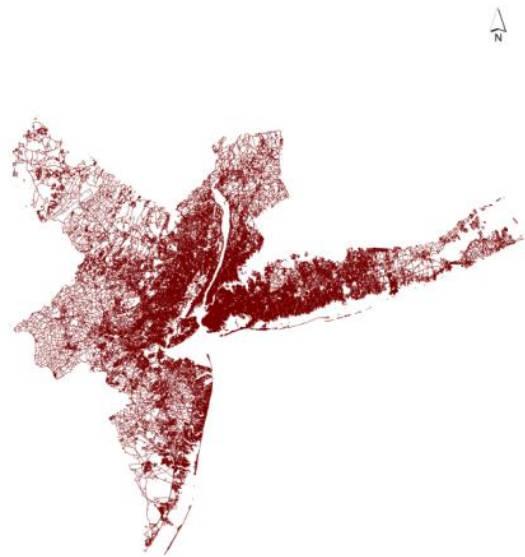
New York, NY

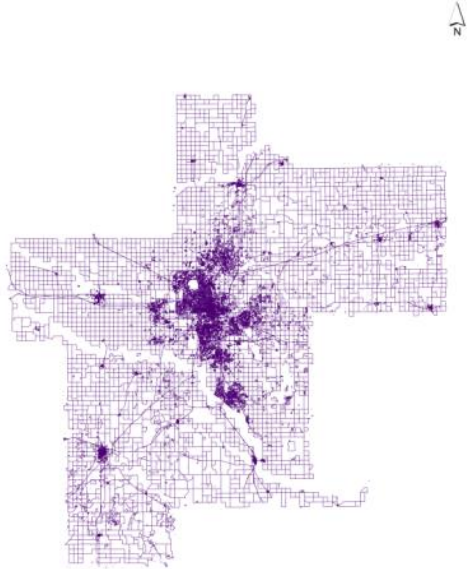
Oklahoma, OK

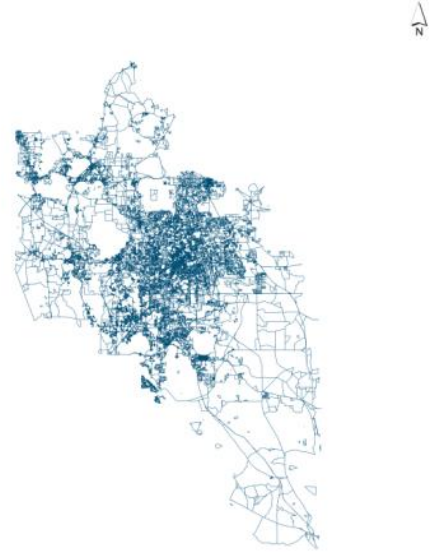
Orlando, FL

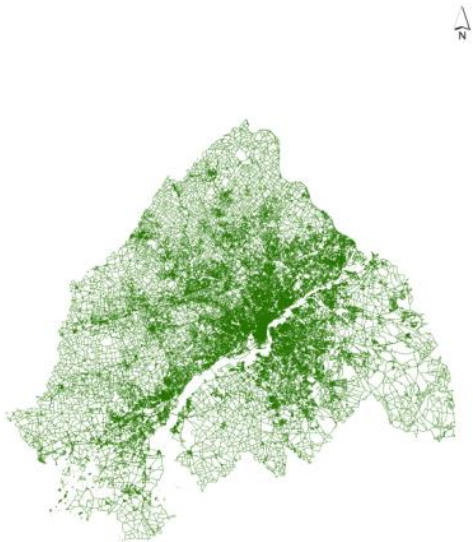
Philadelphia, PA

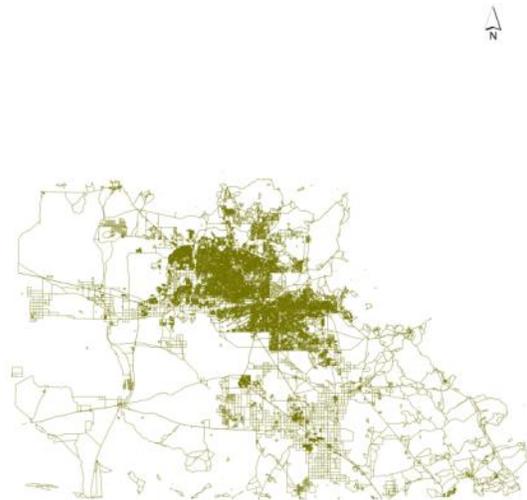
Phoenix, AZ

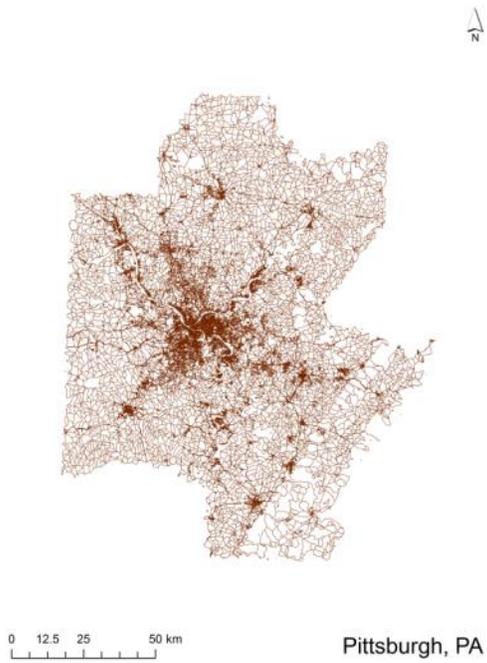
Pittsburgh, PA

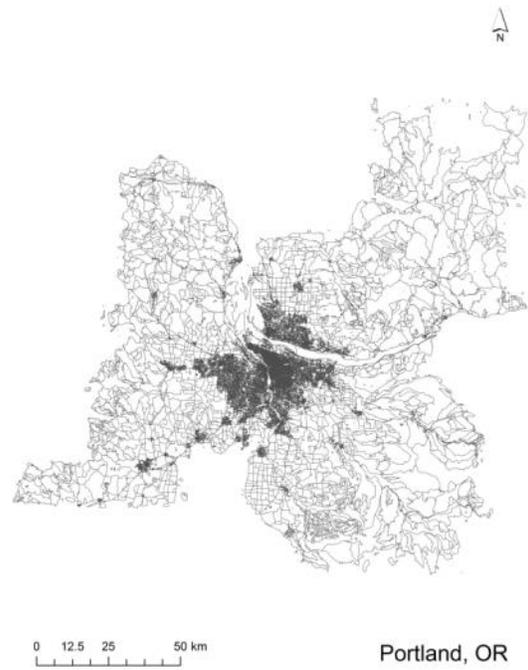
Portland, OR

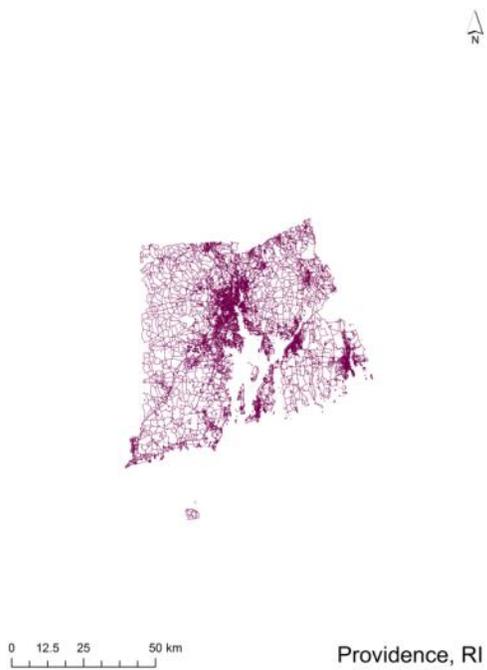
Providence, RI

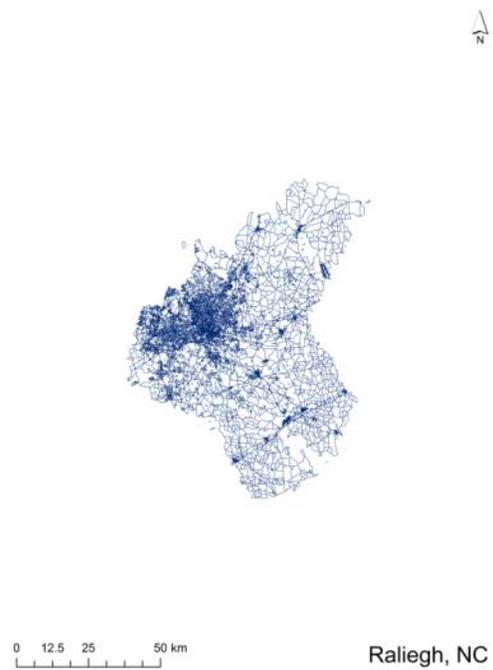
Raliegh, NC

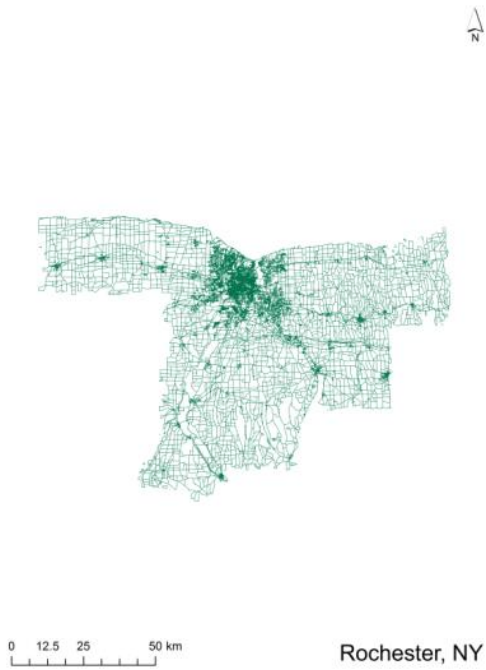
Rochester, NY

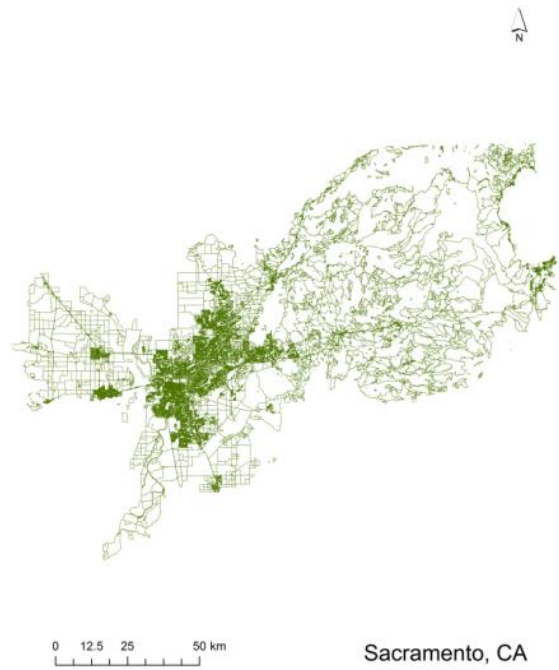
Sacramento, CA

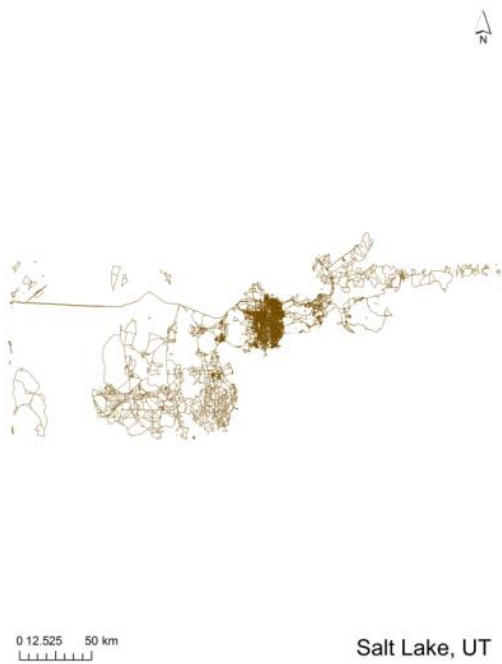
Salt Lake, UT

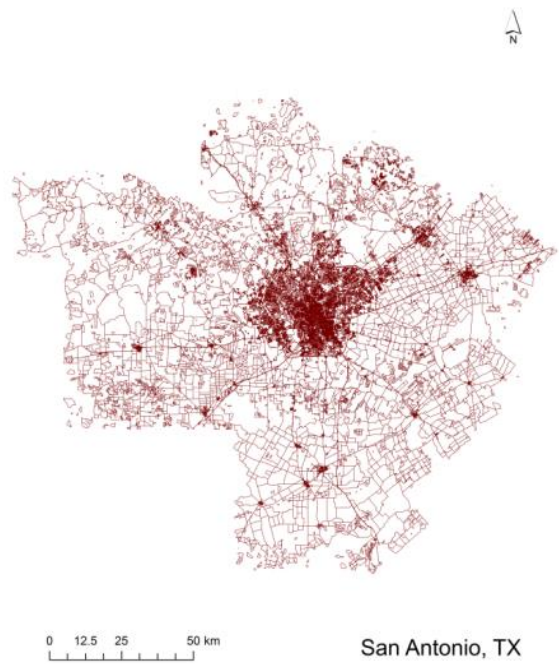
San Antonio, TX

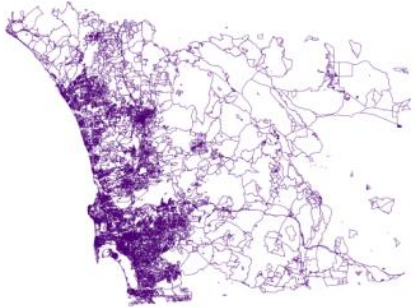
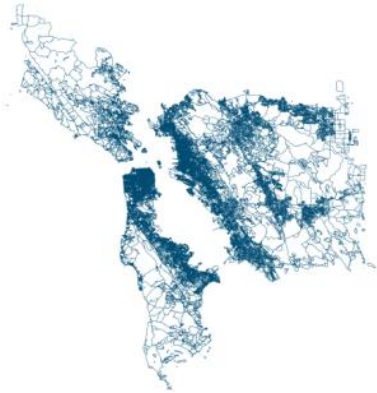
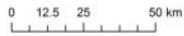
San Diego, CA
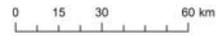
San Francisco, CA

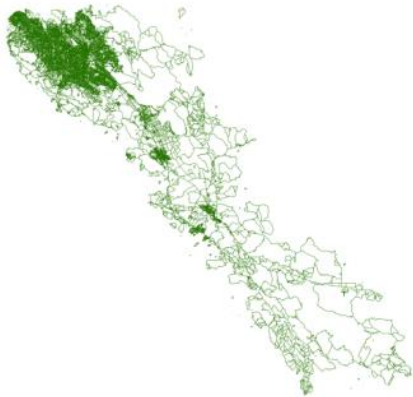
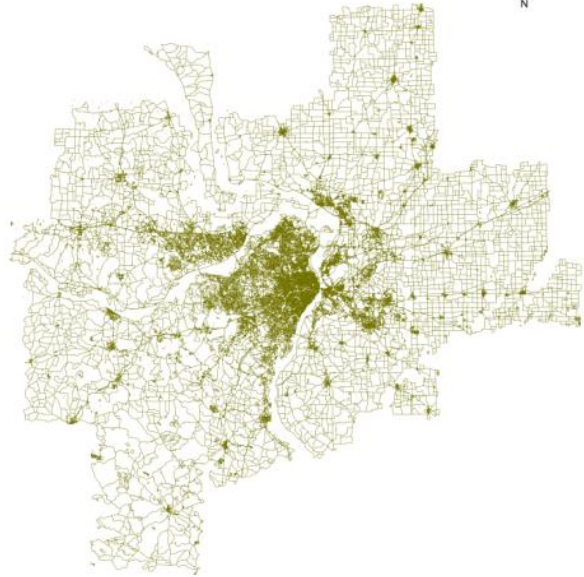
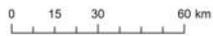
San Jose, CA
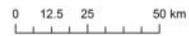
St. Louis, MO

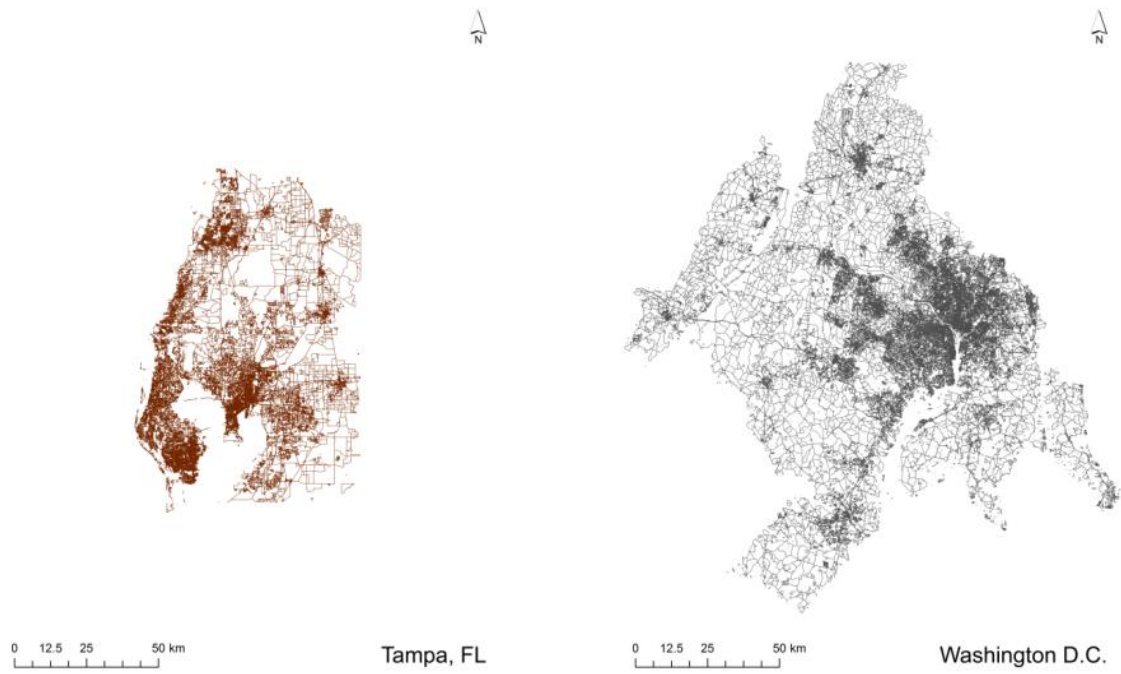

Figure D. Relative sizes and shapes of the 50 urban road polygons analyzed in this study.

E. Discussion of area threshold

As we mentioned before, 'older' cities with lower area thresholds tend to have smaller blocks and are hence more walkable, while in comparison 'younger' cities in general tend to have larger block sizes and thus higher area thresholds, and hence are less walkable. Figure E below presents a comparative same-scale illustration of the road polygons in Phoenix, AZ versus Chicago, IL that have the largest and medium area thresholds, respectively. We can see that in general Chicago offers a more inviting environment towards walking than Phoenix.

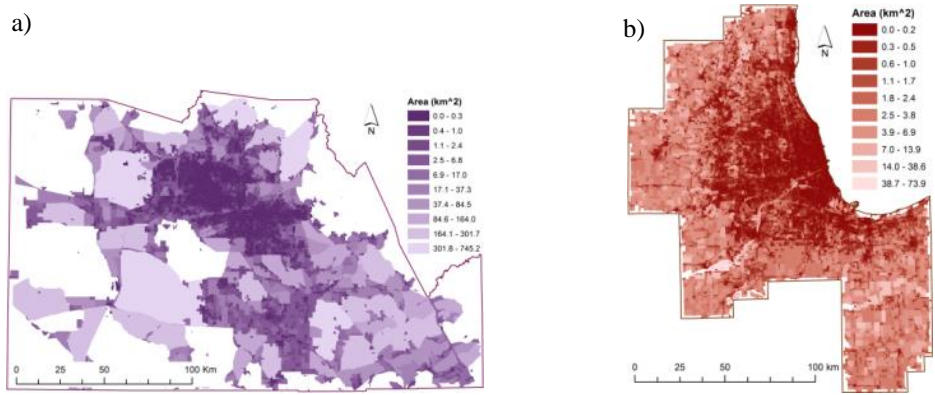

Figure E. a) Phoenix, and b) Chicago road polygons variations

F. Discussion of line threshold

Figure F compares the length variations within the road networks of Salt Lake City, UT and Chicago, IL that have the largest and medium line thresholds, respectively. This is an example of the inland Salt Lake City versus a logistic hub coastal city like Chicago that is a center of freight activity. We see that as a result, Chicago's road network is more compact and has a more uniform distribution of road segments as compared to the non-coastal Salt Lake City that has larger variation in its road segment lengths.

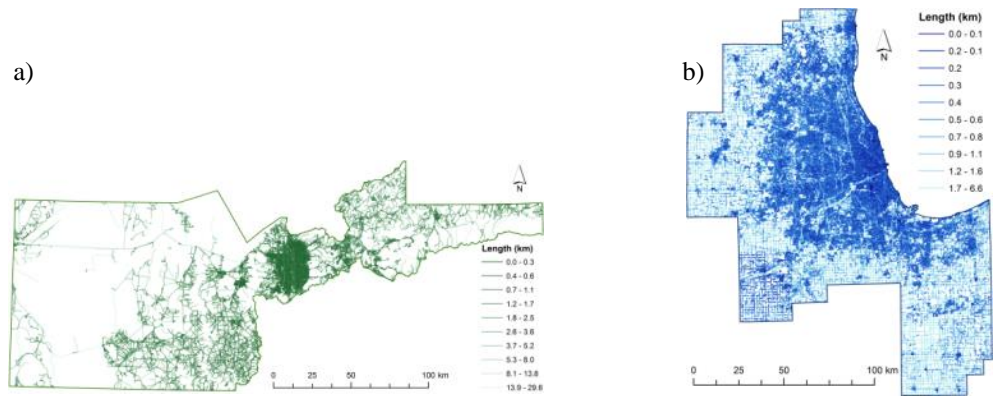

Figure F. a) Salt Lake City, and b) Chicago road length variations

G. Area, point, and line thresholds for 50 U.S. urban road systems

| Urban Area, State | Area Threshold (m) | Line Threshold (m) | Point Threshold (m) |
|---|---|---|---|
| Atlanta, GA | 872 | 610 | 270 |
| Austin, TX | 794 | 666 | 272 |
| Baltimore, MD | 390 | 352 | 153 |
| Boston, MA | 480 | 353 | 174 |
| Buffalo, NY | 971 | 615 | 270 |
| Carson, NV | 156 | 778 | 279 |
| Charlotte, NC | 672 | 627 | 267 |
| Chicago, IL | 984 | 321 | 179 |
| Cincinnati, OH | 745 | 668 | 249 |
| Cleveland, OH | 821 | 479 | 251 |
| Columbus, OH | 907 | 722 | 267 |
| Dallas, TX | 971 | 472 | 215 |
| Denver, CO | 1671 | 654 | 241 |
| Detroit, MI | 751 | 381 | 204 |
| Grand Rapids, MI | 792 | 926 | 395 |
| Hartford, CT | 545 | 514 | 245 |
| Honolulu, HI | 454 | 361 | 178 |
| Houston, TX | 904 | 450 | 210 |
| Indianapolis, IN | 863 | 575 | 213 |
| Jacksonville, FL | 923 | 670 | 271 |
| Kansas City, KS | 1028 | 793 | 282 |
| Las Vegas, NV | 1330 | 484 | 181 |
| Lewiston, ID | 663 | 980 | 727 |
| Los Angeles, CA | 962 | 230 | 152 |
| Louisville, KY | 767 | 879 | 327 |

| Urban Area, State | Area Threshold (m) | Line Threshold (m) | Point Threshold (m) |
|---|---:|---:|---:|
| Memphis, TN | 960 | 891 | 348 |
| Miami, FL | 1660 | 248 | 174 |
| Milwaukee, WI | 700 | 386 | 212 |
| Minneapolis, MN | 904 | 502 | 207 |
| Nashville, TN | 868 | 919 | 383 |
| New Orleans, LA | 699 | 372 | 189 |
| New York, NY | 501 | 282 | 170 |
| Oklahoma, OK | 955 | 828 | 296 |
| Orlando, FL | 1374 | 418 | 200 |
| Philadelphia, PA | 648 | 378 | 197 |
| Phoenix, AZ | 1200 | 535 | 221 |
| Pittsburgh, PA | 707 | 596 | 267 |
| Portland, OR | 787 | 722 | 270 |
| Providence, RI | 531 | 444 | 201 |
| Raliegh, NC | 637 | 562 | 231 |
| Rochester, NY | 881 | 874 | 380 |
| Sacramento, CA | 821 | 627 | 260 |
| Salt Lake, UT | 1033 | 957 | 397 |
| San Antonio, TX | 875 | 806 | 347 |
| San Diego, CA | 1129 | 413 | 186 |
| San Francisco, CA | 640 | 272 | 155 |
| San Jose, CA | 773 | 478 | 195 |
| St. Louis, MO | 880 | 753 | 287 |
| Tampa, FL | 756 | 315 | 180 |
| Washington D.C. | 467 | 361 | 162 |